# Period Variations of Delta Scuti Stars

Michel Breger

*Department of Astronomy, University of Vienna, Türkenschanzstr. 17, A-1180 Wien, Austria*
*and*
*Department of Astronomy, University of Texas, Austin, TX 78712, USA*

**Abstract.** In most Delta Scuti stars, the measured period changes are considerably larger than those expected from stellar evolution. In order to study these period and amplitude changes, a few selected stars are measured photometrically for hundreds of nights with a dedicated telescope. These measurements cover several years or decades. The Delta Scuti stars provide more information (than classical pulsators) since a number of simultaneously excited radial and nonradial modes can be studied. The present results indicate the presence of at least two effects: beating of independent modes with close frequencies and stellar cycles. For period and amplitude changes with time scales less than one year, we confirm the beating hypothesis in three stars. This is shown by the correctly correlated relationship between amplitude and phase changes as well as the repetitions of these cycles. However, the observed period variations with longer time scales are not due to simple beating between two close frequencies. For the star 4 CVn we can derive accurate annual frequency values for at least seven radial and nonradial modes. The annual phases are in excellent agreement with predictions from nearby years, thereby confirming the values and their observed long-term changes. For prograde and retrograde modes, the period variations are of identical size, but with opposite signs. The radial mode shows no (or little) changes. Furthermore, all period variations show a reversal around 1990. These results suggest long-term, regular cycles affecting individual modes differently with some common systematic behavior.

## INTRODUCTION

The periods and amplitudes of most types of pulsating stars are not stable. Some of the causes of these variations are understood, e.g., stellar evolutionary changes and light-time changes in multiple stellar systems. However, in the majority of types of pulsators the observed changes require other or additional explanations. Examples are provided by the Delta Scuti stars (Breger & Pamyatnykh 1998).

The timescales and effects of the modulations of the periods and amplitudes look different in different types of pulsators. Consequently, the studies generally concentrate on individual groups. This may not be the optimum strategy, since the same (presently unknown) astrophysical process may operate in these stars. This process may lead to different symptons in different stars, all of which provide clues. At this conference, we discussed the variations of the cepheid Polaris, the RR Lyrae stars (where the effect is called the Blazkho Effect) and Delta Scuti stars (the present paper). Delta Scuti stars are more difficult to study due to their many simultaneously excited modes. They may provide valuable insight into how the unknown process affects different radial and nonradial modes.

In studying phase shifts, it is essential to examine and eliminate the effects of binary motion, since the varying light-travel time of stars in binary systems leads to regular phase shifts. An example is the Delta Scuti star SZ Lyn (Derekas et al. 2003). However, most observed phase variations cannot be caused by binary motion, since different modes in the same star show different (O-C) phase shifts (e.g., in AE UMa, Szeidl 2001) and since amplitude variations accompany the phase shifts. Could the variations be explained by the beating of close frequencies, possibly a radial and a nonradial mode with almost the same frequency or do we need an exotic process inside the star?

One of the projects at the Vienna Asteroseismology Center consists of long-term monitoring of selected Delta Scuti stars with up to six months of high-precision photometry per year. This makes it possible to detect between 50 and 100 pulsation modes with excellent frequency resolution, obtain mode identifications for 10 to 20 of these modes, and to provide the observational data for comparison with stellar models. An extremely valuable byproduct of these campaigns is the possibility to look at the amplitude and period variability in detail. The present paper presents preliminary results for the star 4 CVn, for which hundreds of nights obtained with the 75-cm Vienna Automatic Telescope in Arizona (Breger & Hiesberger 1999) have been analyzed.

## TIME SCALES UNDER ONE YEAR: CLOSE FREQUENCIES

We have previously found that in the well-studied stars FG Vir and BI CMi, the amplitude and period variations with timescales less than 250d can be fit very well by a model of beating between two close frequencies. The observed amplitude and phase variations of a number of frequencies with amplitude and phase variability are exactly as predicted from a two-frequency solution; the agreement even repeats from beat cycle to beat cycle. Examples are FG Vir and BI CMi (Breger & Bischof 2002, Breger & Pamyatnykh 2006).

In fact, the present study of 4 CVn adds another example: the amplitude and phase variations of the 6.12 cycle $d^{-1}$ frequency can be fit perfectly by a close, but observationally well separated, doublet: 6.1170 and 6.1077 cycle $d^{-1}$ with a beat period of 107 d. In the power spectrum this shows up as a clean doublet.

Several years ago, the large number of detected close-frequency pairs seemed difficult to explain if one assumes that the frequencies of $\ell = 1$ and 2 modes are randomly distributed. Of course, if we also include modes with higher $\ell$ values, the number of possible accidental agreements is increased and no problem of interpretation would exist. However, high-$\ell$ modes have small photometric amplitudes (averaged over the stellar disk) and would therefore not be able to produce the strong beating. For the close-frequency hypothesis to be valid, either the $\ell > 2$ modes have artificially increased amplitudes (e.g., through resonance) or a selection mechanism exists to select nonradial frequencies near the radial frequencies. We *now* know that in Delta Scuti stars radial and nonradial modes tend to cluster around the radial modes due to trapped modes in the envelope (Breger, Lenz & Pamyatnykh 2009). This, in turn, can at least partially explain the large number of close frequencies in Delta Scuti stars.

## TIME SCALES OF YEARS: STELLAR CYCLES IN 4 CVN

For modulation time scales of several years or decades, the observational test for the close-frequency hypothesis is difficult. The reason is the lack of near-continuous coverage due to the presence of long observing gaps from year to year as well as the difficulty of observing a particular star every year. Consequently, special observing campaigns are required to examine such long-period modulations.

### The Observational Campaign Of 4 CVn

From 2005 to 2009, the Vienna Automatic Photoelectric Telescope (APT) in Mount Washington, Arizona, USA was utilized for five months per year to measure 4 CVn relative to two comparison stars. The preliminary multifrequency analysis with PERIOD04 (Lenz & Breger 2005) led to the determination of 72 significant frequencies, of which 60 had small amplitudes of 0.001 mag or less. In order to study the amplitude and period variations of the modes with relatively large amplitudes on a near monthly basis, the number of degrees of freedom need to be reduced: consequently, the best 'average' solution for frequencies f13 to f72 was prewhitened. Subsequent simulations revealed that that the effects of such prewhitening on the subsequent analysis was small.

### Results For 4 CVn

Figure 1 shows the frequency and amplitude variations for one of the 'large-amplitude' frequencies from 2005 to 2009. Here we have used the standard (O-C) approach in which a best frequency is adopted and the phase shifts for different time spans are calculated. The figure demonstrates that the amplitudes and frequencies are variable on a timescale of many years or even decades. Modulations with time scales under one year are not evident. Other pulsation modes examined confirm the results (except for the modulation produced by the close-frequency pair near 6.11 cycle $d^{-1}$ mentioned earlier).

In order to examine the properties of the long-term changes in more detail, we have analyzed the collected data from 1966. The (O-C) plots revealed a common problem with the phase shift method: the unknown number of cycles between the different years. For the star 4 CVn, the (O-C) method is not optimum because of the sign reversals of dP/dt around 1990. We will therefore no longer assume an average period (and calculate phase shifts), but derive annual (or two-year) period values independent of the masurements from other years.

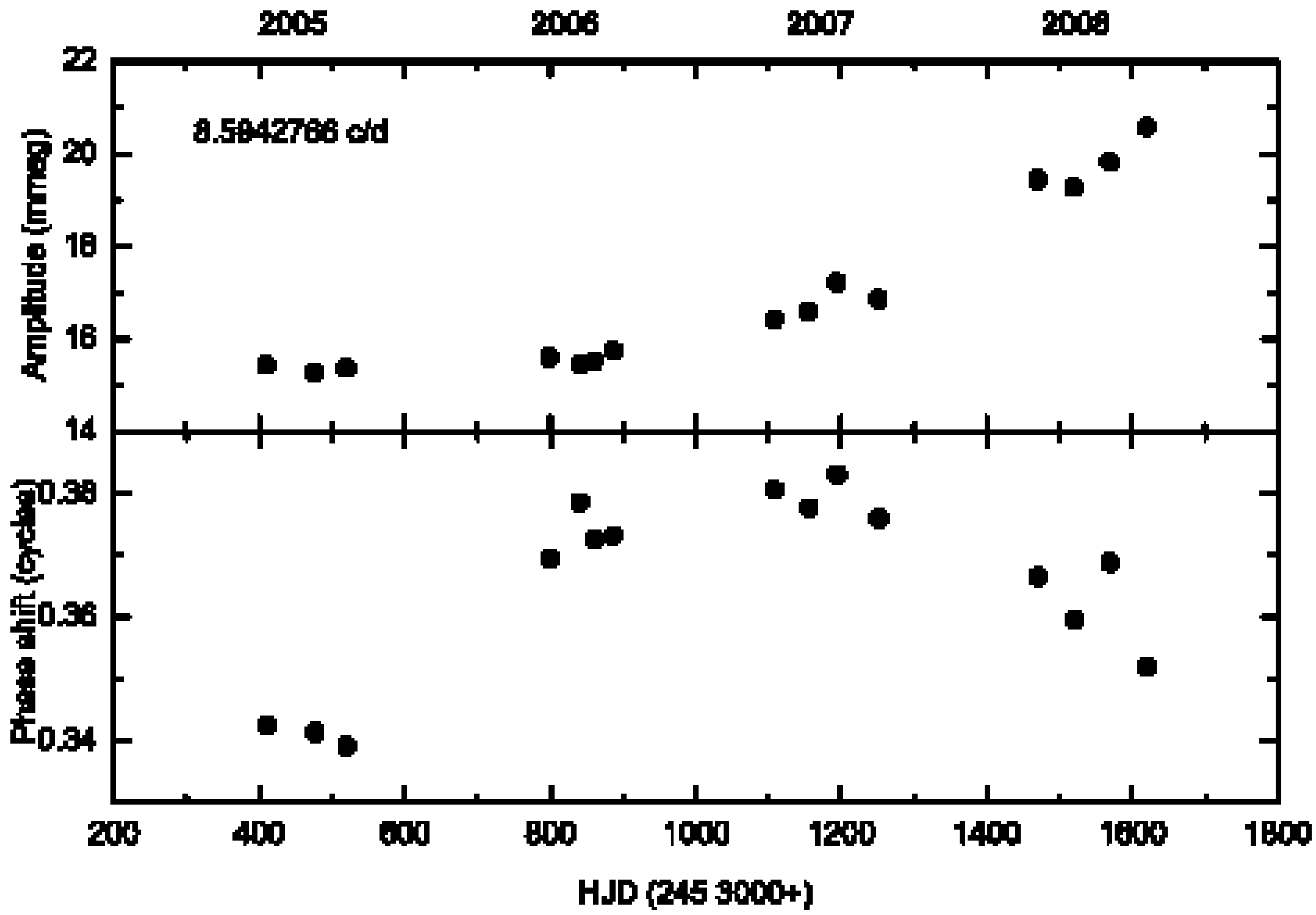

**FIGURE 1.** Amplitude and phase variations of the 8.59 cycle $d^{-1}$ nonradial mode in 4 CVn from 2005 to 2008. The phase shifts were calculated using the frequency value listed at the top left: due to the variable frequency the last few digits are essentially meaningless. The diagram shows that both the period length and the amplitude of the mode were increasing. The observed changes within each year match the overall behavior from 2005 through 2008, suggesting changes with time scales of years or decades.

Fortunately, even for a few data sets prior to the 2005 - 2009 time period the available data are sufficient to reliably derive the annual (or two-year) period values for the modes with high amplitudes. These data were the extensive Fitch measurements from 1974, and the multisite Delta Scuti Network campaigns from 1983/4 and 1996/7 (see Breger 2000 for details on these as well as smaller data sets starting in 1966). Furthermore, observations from 1991/1992 have also recently become available (Breger, Davies & Dukes 2008).

Figure 2 shows the derived period changes for three pulsation modes with different values of the azimuthal quantum number *m*, taken from Castanheira et al. (2008). Although we present here only the results for three modes, the other pulsation modes behave in a similar manner. In particular, we note that for the longer modulation periods examined here:

(i) the radial mode only show small or zero period changes. However, the amplitudes are variable (not shown),

(ii) the nonradial $\ell$ = 1 mode is very unstable: between 1997 and 2007 a period change of (1/P) dP/dt ~ ± 6.7 x 10 $^{-6}$ yr $^{-1}$ is found. This is about two orders of magnitude than the value expected from stellar evolution,

(iii) the sign of dP/dt depends on the *m* value,

(iv) there was a sign reversal of dP/dt around 1990. This is seen in the modes with positive as well as the modes with negative dP/dt changes,

(v) we have not seen any abrupt period breaks similar to those reported for V1162 Ori by Hintz et al. (1998). For an excellent summary on the V1162 Ori we refer to Arentoft & Sterken (2002),

(v) the amplitude changes (not shown here) are not synchronized with the period changes. So far, we have not been able to find any relation between the two.

The observed long-term behavior is in disagreement with the close-frequency as well as evolution hypotheses and supports the existence of stellar cycles. The reason for these cycles is presently unclear.

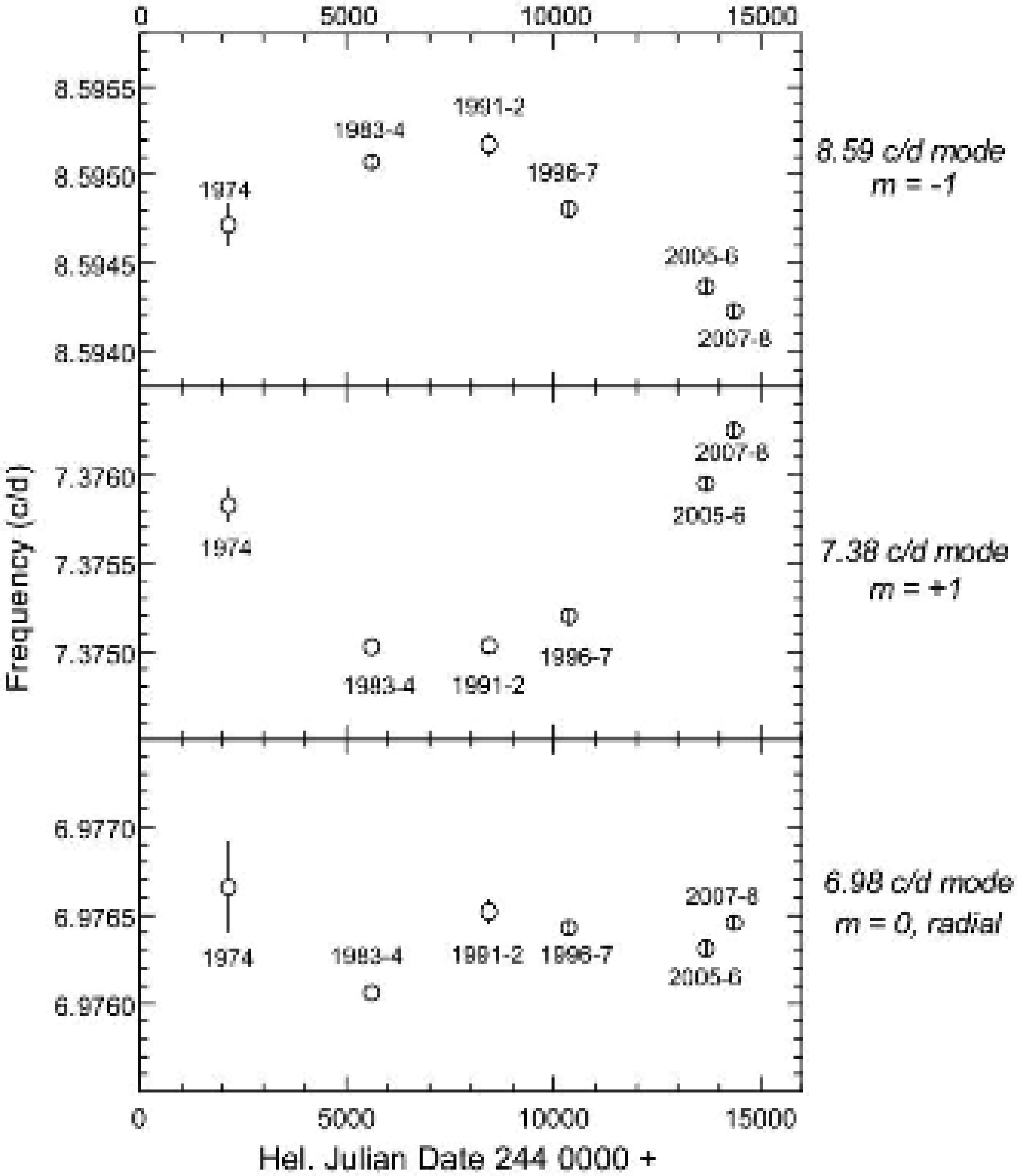

**FIGURE 2.** Frequency variation of three modes with different azimuthal numbers. The formally derived uncertainties are shown. Note the similarity in the sizes and phasing of the frequency changes in the $m$ = +1 and -1 modes as well as the different signs of dP/dt. Some (but not all) other nonradial modes show a similar behavior.

## ACKNOWLEDGMENTS

It is a pleasure the acknowledge helpful discussions with Patrick Lenz and Alosha Pamyatnykh. This investigation has been supported by the Austrian Fonds zur Förderung der wissenschaftlichen Forschung.